# 1 Bit Electronically Reconfigurable Transmitarray Antenna With Out-of-Band Scatter Suppression

Binchao Zhang, Fan Yang, Shenheng Xu, Maokun Li, and Cheng Jin

*Abstract*- Stealthy electronically reconfigurable transmitarray antennas are essential components in wireless communication and radar detection systems. Therefore, this paper proposes a 1 bit electronically reconfigurable transmitarray antenna with out-of-band scatter suppression. The transmitarray consists of two layers, the absorptive frequency selective transmission (AFST) layer and reconfigurable transmitarray (RTA) layer, separated by air. Specifically, the AFST layer achieves out-of-band scattering suppression and in-band transmission performance by utilizing the first three resonant modes of a bent metallic strip with a centrally loaded resistor. Additionally, the RTA layer adopts a receiver-transmitter structure with an active receiving dipole and a passive orthogonal transmitting dipole. The 1 bit phase shift is achieved by alternating two pin diodes integrated on the active dipole to reverse its current direction. To evaluate the proposed design, a 16×16-element transmitarray prototype is designed, fabricated and measured. For scattering, the 10-dB radar cross section reduction is realized within 4~5.2 GHz and 10.9~11.4 GHz, respectively. For radiation, the measured gain is 19.9 dBi at 7.5 GHz, corresponding to an aperture efficiency of 12.1%. and the beam scanning covers $\pm 60^o$ with gain loss of 5 dB in both two principal planes.

## I INTRODUCTION

In modern long-distance communication and radar systems, high-gain antennas are essential for achieving a desirable signal-to-noise ratio. Over the past few decades, a variety of array technologies have been developed at a rapid pace, including phased array antennas, reflectarray antennas, and transmitarray antennas [1]-[4]. However, high-gain antennas often have a large aperture, resulting in a strong scattering field that increases the risk of enemy radar detection [5]-[6]. To address this issue, it is necessary to design a high-gain antenna that also exhibits low scattering performance.

Reflectarrays and transmitarrays have gained popularity as competitive high-gain antenna candidates due to their ability to achieve high gains without the use of complex and lossy feed networks. Compared to reflectarrays, transmitarrays do not suffer from feed blockage, leading to substantial advancements in their design and implementation [7]-[10]. However, achieving Radar Cross Section (RCS) reduction for transmitarray antennas can be challenging. This is because it is difficult to reduce out-of-band RCS while maintaining low in-band insertion loss transmission and 360° phase-modulation capability.

Various approaches have been explored to realize low RCS transmitarray antennas. These include the use of a resistive sheet, fixed bandstop Frequency Selective Surface (FSS), and adjustable bandpass FSS with phase-shifting capability to achieve absorption-type low RCS transmitarray antennas [11]. Building on this, a diffuse reflection low RCS transmitarray antenna was realized by using anti-reflection dielectric columns of different heights and a bandpass FSS as a phase shifter [12]. In addition, an asymmetric resonator with resistance was used to obtain absorption-transmission-absorption response, and the transmission phase was controlled by rotating an open resonant ring to achieve a 1-bit phase change [13]. The same design concept of integrating an absorptive frequency selective transmission (AFST) layer and phase modulation layer was also implemented in [14]-[15].

Compared to traditional transmitarray antennas, 1 bit reconfigurable transmitarray antennas (RTAs) offer several advantages, including low cost and electrically controllable radiation characteristics. As a result, they are better suited for modern high-integration application systems that require various functions. Previous research on 1 bit reconfigurable transmitarray antennas has primarily focused on improving bandwidth, polarization, and scanning angle performance [16]-[25]. However, there has been limited research on reducing the RCS of transmitarray antennas. In [26], 1 bit beam scanning and RCS reduction were achieved in different polarizations, but the structures were relatively complex. Thus, reducing co-polarized RCS for reconfigurable transmitarray antennas without negatively affecting their radiation performance remains a challenging problem to solve.

To address this technology gap, we propose a 1 bit electronically reconfigurable transmitarray with out-of-band scatter suppression. The transmitarray comprises an AFST layer and an RTA layer. The AFST layer is utilized to suppress out-of-band scattering while maintaining in-band transmission performance, and the RTA layer is responsible for achieving 1 bit transmission phase shifting. To verify the design concept, we designed, fabricated, and measured a 16x16-element transmitarray prototype. The measured radiation and scattering results demonstrate that the proposed low RCS RTA achieves high-gain in-band radiation and beam scanning while effectively suppressing out-of-band scattering.

This work is supported in part by the National Key Research and Development Program of China under Grant No. 2020YFB1806302 and in part by THE XPLORER PRIZE. (*Corresponding author: Fan Yang*)

B. Zhang, F. Yang, S. Xu, and M. Li are with the Department of Electronic Engineering, Tsinghua University, Beijing 100084, China. (Email: zhangbinchao@tsinghua.edu.cn)

C. Jin is with the school of Cyberspace Science and Technology, Beijing Institute of Technology, Beijing, 100081, China.

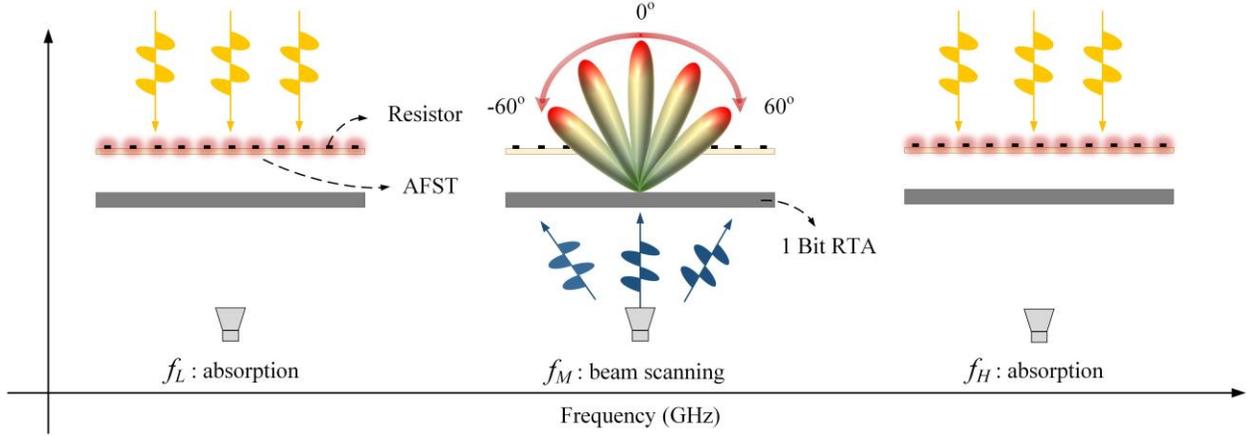

Fig. 1 Design schematic diagram of the proposed low RCS 1 bit RTA with various features at different frequencies, (a) $f_L$, (b) $f_M$, and (c) $f_H$.

## II DESIGN CONCEPT

The key step in designing the low RCS 1 bit RTA is to construct two RCS reduction bands in the lower and upper sides of the antenna's operating band while maintaining high-gain radiation and beam scanning in the middle band. The entire structure consists of two layers separated by an air cavity: the AFST and RTA layers, with the feed antenna located under the RTA layer at a specific distance. The designed low RCS RTA exhibits various frequency responses in different frequency bands, and its different components offer different functionalities, as described in Fig. 1. Each component of the integrated structure must work in cooperation with one another to create three artificial bands with distinct frequency responses.

As shown in Fig. 1, the AFST layer functions in the out-of-band frequencies $f_L$ and $f_H$ by absorbing low and high-frequency incident electromagnetic waves with its embedded resistance. Meanwhile, incident electromagnetic waves of $f_M$ can be transmitted with low insertion loss to ensure in-band high-gain radiation characteristics. In contrast, the 1 bit RTA operates in $f_M$ and plays a crucial role in regulating the phase of the incident electromagnetic wave from the feed horn, thereby achieving high-gain radiation performance and beam scanning. The RTA layer should provide AFST with complete reflection characteristics similar to PEC in $f_L$ and $f_H$ to ensure RCS reduction performance in these two frequency bands.

## III ELEMENT ARCHITECTURE

The design process of the low RCS 1 bit RTA, based on the design concepts outlined in Section II, can be divided into three main steps: AFST design, 1 bit RTA design, and array design. Each of these steps is crucial in achieving the desired low RCS and high-gain and beam scanning radiation performance of the transmitarray antenna.

### A  AFST Element

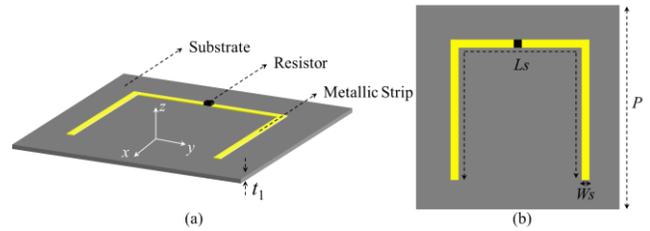

Fig. 2 Element of the proposed AFST in (a) perspective view and (b) top view.

In the first step of the low RCS 1 bit RTA design process, the AFST element was analyzed and designed. The AFST needed to have the ability to out-of-band absorption and in-band transmission while maintaining a low profile and simple structure. To achieve these goals, multi-resonant modes on the metallic strip were used, with different functions assigned to different modes by loading resistor [27]-[28]. The element of the AFST is shown in Fig. 2, where a bent metallic strip is printed on a dielectric substrate with a relative dielectric constant of 3, and a lumped resistor is loaded in the center of the strip. The element has a period $P$ of 20 mm, approximately half wavelength corresponding to the frequency $f_H$, and the length $L_s$ of the metallic strip is 40 mm, roughly equal to one wavelength corresponding to the frequency $f_H$. Other physical parameters are listed: $P = 20$ mm, $t_1 = 0.5$ mm, $L_s = 40$ mm, $W_s = 0.6$ mm, and $R = 250\ \Omega$..

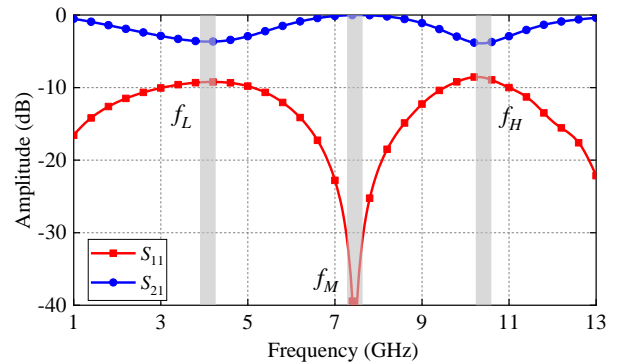

Fig. 3 Simulated S parameters of the proposed AFST.

The designed AFST structure is simulated to verify its ability to out-of-band absorption and in-band transmission perfor-mance. The simulation is carried out using the HFSS commercial simulation software with Floquet ports and Master-Slave boundary conditions. Fig. 3 shows the simulated $S$ parameters, which indicates that the incident electromagnetic wave around $f_L = 4$ GHz and $f_H = 10.5$ GHz is absorbed. In contrast, at $f_M = 7.5$ GHz, the incident electromagnetic wave passes through with low insertion loss. Therefore, the designed AFST structure can be applied to the low RCS 1 bit RTA to achieve out-of-band RCS reduction performance.

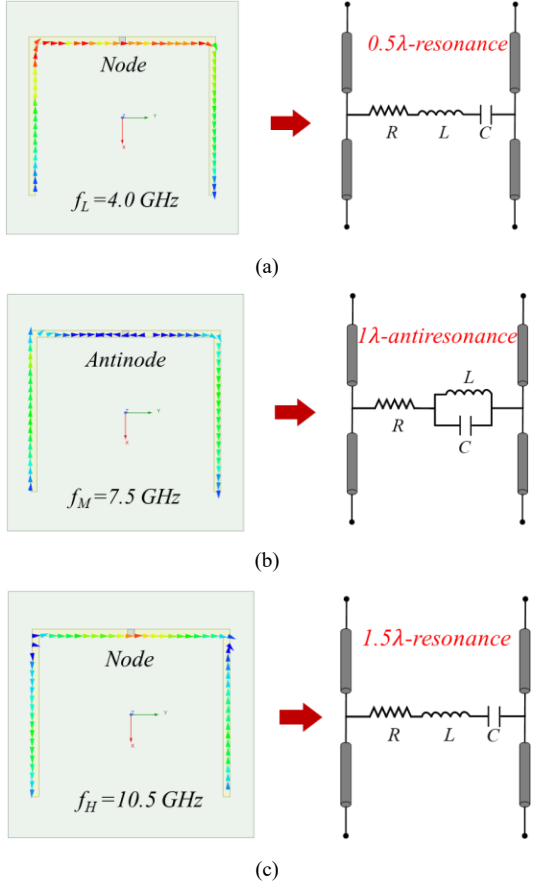

Fig. 4 Simulated current distributions and their corresponding equivalent circuit models at different resonant frequencies, (a) $f_L$, (b) $f_M$ and (c) $f_H$.

The working principle of the AFST's absorption-transmission-absorption can be analyzed by observing its current distribution at the central frequency of $f_L$, $f_M$ and $f_H$, as shown in Fig.4. By observing the current intensity at the position of the resistor, it is found that with the change of the resonant frequency, the current intensity presents a change in the state of node-antinode-node. Specifically, when the resistor is in the current wave node state, the incoming electromagnetic wave will be absorbed, and when the resistor is in the current antinode state, the resistor is ineffective. On the other hand, the working principle of AFST can be further analyzed by equivalent circuit model. The half- and one-and-half-wavelength resonances of the metallic strip can be equivalent to $RLC$ series circuit, with the current reaching its maximum value at resonance. In contrast, for one-wavelength resonance, it is equivalent to an $LC$ parallel circuit in series with $R$, resulting in the current tending to zero at resonance.

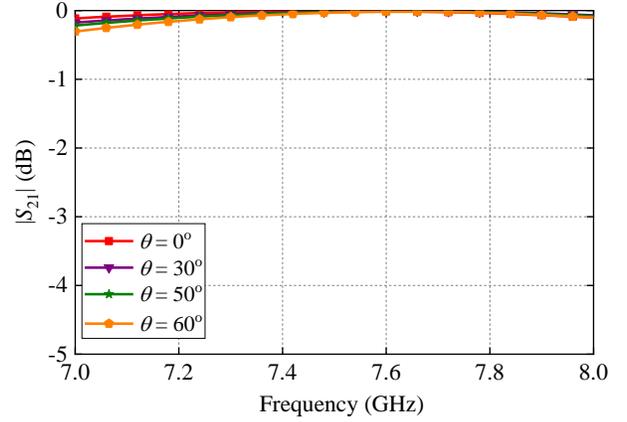

Fig. 5 Simulated transmission coefficients of the proposed AFST under oblique incidences.

In addition, it is important for the AFST to have good in-band transmission performance under oblique incidence to reduce beam scanning gain loss. To evaluate this, the transmission coefficients of the AFST were simulated at different oblique incidence angles, as shown in Fig. 5. The results indicate that the proposed AFST has a stable transmission coefficient for incidence angles up to 60°, ensuring a stable high gain during beam scanning. This performance is crucial for the low RCS 1 bit RTA to maintain its radiation characteristics even under large beam scanning angle, and highlights the effectiveness of the AFST design.

### B  1 Bit RTA Element

After the design and analysis of the AFST, the next step is to design the 1 bit RTA element. As mentioned in Section II, the RTA structure must meet two key requirements. The first is to provide low insertion loss transmission and 1 bit phase regulation at $f_M = 7.5$ GHz to enable in-band high gain and beam scanning performance. The second is to maintain perfect reflection characteristics near $f_L = 4$ GHz and $f_H = 10.5$ GHz to ensure the out-of-band RCS reduction performance of the AFST.

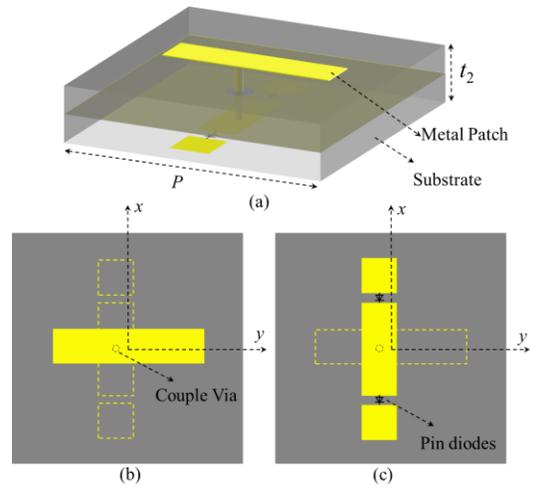

Fig. 6 Element of the proposed 1 bit RTA, (a) perspective view, (b) top view and (c) bottom view.

The proposed 1-bit RTA element is shown in Fig. 6, which employs a receiver-transmitter structure with an active receiving dipole and a passive asymmetric transmitting dipole [29]. Both dipoles, separated by a common ground and connected by a metallized via-hole, are linearly polarized and aligned in the orthogonal direction. Two pin diodes are integrated on the active receiving dipole and biased in opposite states to achieve a 180° phase difference by controlling the conversion direction of line polarization (+x or -x) through the positive and negative voltage of the bias. The ON and OFF states of the pin diode are modeled as *RL* and *RLC* series circuits, respectively [30]. Specifically, in the ON state, the resistance and inductance are $R_{on} = 1\,\Omega$ and $L_{on} = 0.45$ nH, respectively. In the OFF state, the resistance, inductance, and capacitance are $R_{off} = 10\,\Omega$, $L_{off} = 0.45$ nH, and $C_{off} = 0.11$ pF, respectively. The element has a period $P$ of 20 mm, a thickness $t_2$ of 4 mm, and a substrate with a relative dielectric constant of 3.

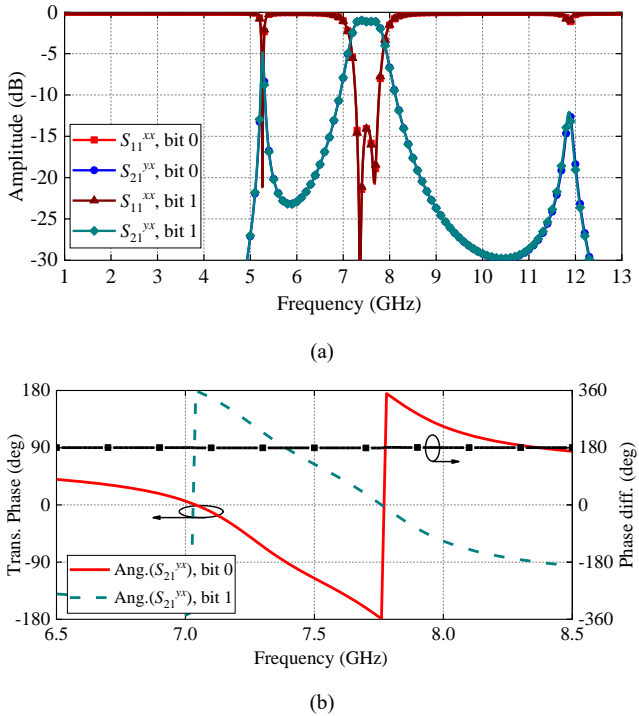

Fig. 7 Simulated S parameters of the 1-bit RTA element under different states, (a) amplitudes, (b) phase.

Fig. 7 illustrates the simulated S-parameters of the designed RTA, including its amplitude and phase. The results indicate that the structure has achieved a good polarization converter transmission performance at 7.5 GHz, with two resonant frequencies in the band. These resonant frequencies are generated by the half-wavelength resonance of the top dipole, which are then differentiated into two frequencies through the coupling of the metallic via-hole. The band insertion loss of the passband is approximately 1.2 dB, which is attributed to the resistance loss of the pin diode. Furthermore, Fig. 7(b) depicts the transmission phase under different states, demonstrating that a 180° phase difference can be constructed within the transmission band through the pin diode, thus achieving a 1-bit electronically RTA. However, it should be noted that the structure has an unexpected resonance point near 5.3 GHz, which may deteriorate the absorption performance of the AFST at this frequency. Therefore, it is necessary to eliminate this resonance frequency.

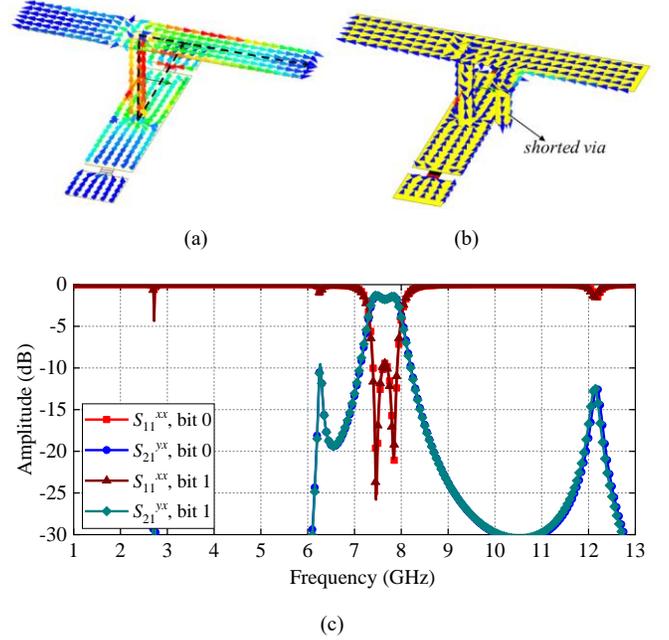

Fig. 8 Simulated current distribution of the RTA at 5.3 GHz (a) without and (b) with shorted via-hole, and (c) the simulated S parameters with shorted via-hole.

By observing the current distribution at the resonant frequency of 5.3 GHz, as shown in Fig. 8(a), a new resonant path was found to be generated through the bottom dipole, via-hole and the top dipole, as indicated by the black dotted line. To eliminate this resonance, a short-circuit metallic via-hole was added to the top dipole to forcibly destroy the resonant path, as shown in Fig. 8(b). It can be seen that the current distribution is destroyed after the addition of the short-circuit via-hole, thus eliminating the polarization converter transmission performance at this frequency. Fig. 8(c) shows the simulated S parameter after adding the short-circuit via-hole. It can be observed that the resonant frequency at 5.3 GHz is eliminated. Although a new resonance appears at 6.25 GHz, the co-polarization reflection coefficient at this point is greater than -1dB, so it has little influence on the out-of-band absorption performance. Notably, the short-circuit via-hole can also be reused as a ground through-hole for the feed voltage, providing a 0 V voltage to the pin diode.

### C  Low RCS 1 Bit RTA Element

The element of the low RCS 1 bit RTA is constructed by integrating the AFST and 1bit RTA structures according to the design concepts in Section II, as shown in Fig. 9(a). The AFST and RTA layers are separated by an air layer with a thickness $h$ of 6.5 mm. Meanwhile, the feed lines of the RTA are given in Fig. 9(b). The influence of feed lines on transmission performance is reduced by polarization isolation. For each receiver-transmitter structure, the middle portion of the active receiving dipole is grounded through the passive transmitter dipole, while a +2.5/-2.5 V DC biasing voltage is applied to both ends of the active receiving dipole to control the pin diode states. The quarter-wavelength transformer and open-ended radial stubs on the receiving layer are designed to choke the RF leakage. The biasing point is positioned close to the zero point of the electric field at the active receiving dipole to minimize the influence of the bias lines on the RF performance.

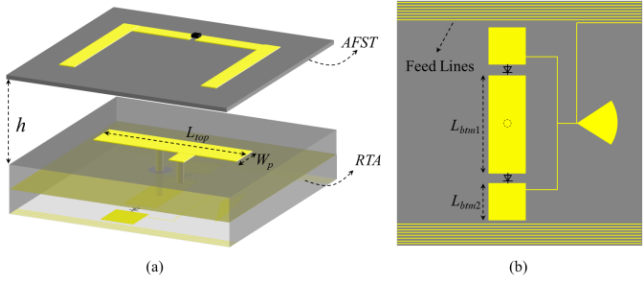

Fig. 9 Element of the proposed low RCS 1 bit RTA, (a) perspective view, (b) bottom view.

Next, in order to better improve the out-of-band stealth performance of RTA, the structural parameters of AFST need to be further optimized. The main parameters involved are the width $W_s$ of the metallic strip and the loaded resistance value $R$. The required parameter combination can be found through the variation trend of the amplitude value of the lowest point in the low and high absorption bands with $W_s$ and $R$, as shown in Fig. 10. By observing Fig. 10(a), it is seen that the absorption performance of the low frequency band gradually increases with the increase of $W_s$ and $R$. However, it can be found from Fig. 10(b) that excessive $W_s$ and $R$ will cause the deterioration of the absorption performance at high frequency. Considering the absorption performance at low and high frequencies, $W_s = 2.5$ mm and $R = 250\Omega$ can be selected. In addition, the remaining parameters of the structure are listed as follows: $L_{top} = 12.2$ mm, $W_p = 2$ mm, $L_{btm1} = 6$ mm, $L_{btm1} = 3$ mm. The diameter of the coupling metallic via-hole is $D_{via} = 0.4$ mm. Its position is 1.2 mm away from the center of the top dipole and located in the center of the bottom dipole.

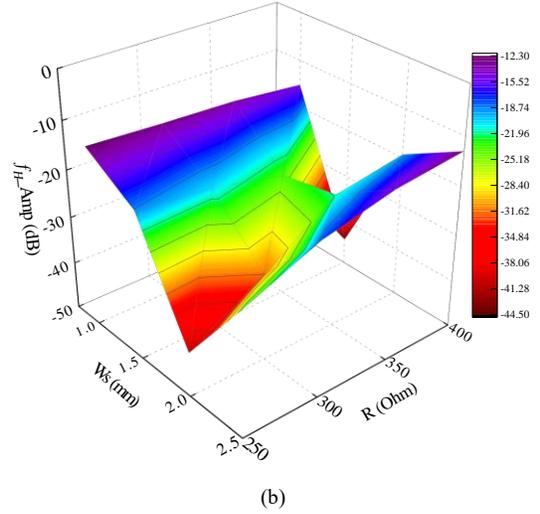

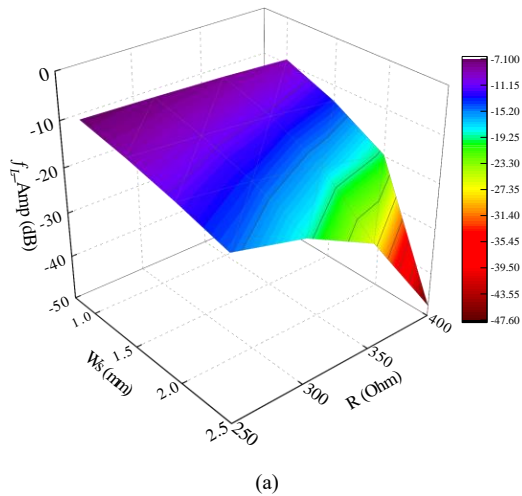

Fig. 10 The absorption properties of (a) low and (b) high frequency bands vary with parameters.

Based on the optimized parameter combination, the element of the proposed low RCS 1-bit RTA was simulated using the HFSS simulation software, and the amplitudes and phases are illustrated in Fig. 11. As shown, the designed structure successfully achieves a 10 dB reduction in RCS in the low frequency range of 3.83-5.41 GHz and high frequency range of 10.82-11.7 GHz. Additionally, it exhibits a 3 dB transmission performance and 1 bit phase reconfiguration in the frequency range of 7.24-7.84 GHz. Consequently, the design is capable of suppressing out-of-band scattering and achieving in-band high gain radiation and beam scanning performance.

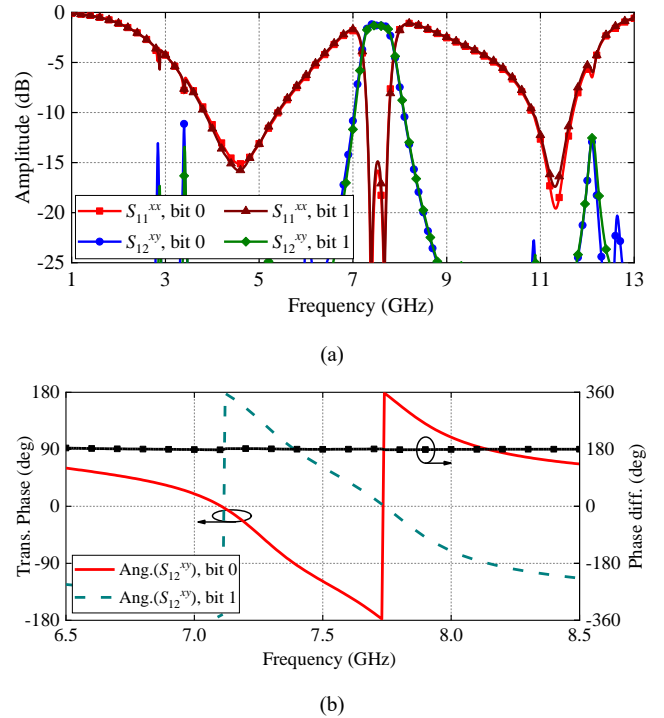

Fig. 11 Simulated S parameters of the low RCS 1-bit RTA unit under different state, (a) amplitudes, (b) phase.

## IV EXPERIMENTAL VERIFICATION

### A    Prototype and Fabrication

A 16×16 low RCS 1-bit RTA prototype is designed, fabricated, and measured based on the element structure described in Section III, as illustrated in Fig. 12. The prototype has an effective aperture of 320 mm×320 mm and integrates 512 pin diodes (SMP1340-040LF). Additionally, there are 4×64 pin connectors located on the sides of the array, which are connected to the FPGA control board to enable electronic control of each element's state. The linearly polarized horn antenna is used as the feed to excite the RTA, with a focal-diameter ratio *F/D* of 0.8. In the case of beam scanning, the actual compensation phase $\Delta\varphi_{mn}$ for each element $(m,n)$ in different beam scanning angle $(\theta_t, \varphi_t)$ can be calculated using the following formula

$$\Delta\varphi_{mn} = \Delta\varphi_f + \Delta\varphi_t + \Delta\varphi \qquad (1)$$

$$\Delta\varphi_f = \frac{2\pi}{\lambda}\left(\sqrt{F^2 + (mP)^2 + (nP)^2} - F\right) \qquad (2)$$

$$\Delta\varphi_t = -\frac{2\pi}{\lambda}(mP\sin\theta_t\cos\varphi_t + nP\sin\theta_t\sin\varphi_t) \qquad (3)$$

Where $\Delta\varphi_f$ and $\Delta\varphi_t$ are the phase difference because of element position and beam scanning, and $\Delta\varphi$ is the reference phase. Then, the bit state of each element can be obtained using the formula (4)

$$bit_{mn} = \begin{cases} 0 & \Delta\varphi_{mn} \in [0,\pi] \pm 2n\pi \\ 1 & \Delta\varphi_{mn} \in [\pi,2\pi] \pm 2n\pi \end{cases} \qquad (4)$$

The measurement process consists of two parts: RCS reduction evaluation and radiation performance measurement. Therefore, two different measurement systems are set up to complete the measurement process, as illustrated in Fig. 12. The system shown in Fig. 12(a) has the prototype in an anechoic chamber that is irradiated by a pair of linearly polarized horn antennas. This system is used to obtain the reflection coefficients and calculate the value of RCS reduction. On the other hand, the fabricated RTA is placed in a near-field anechoic chamber and acted as a transmitting antenna to measure its beam scanning performance, as shown in Fig. 12(b).

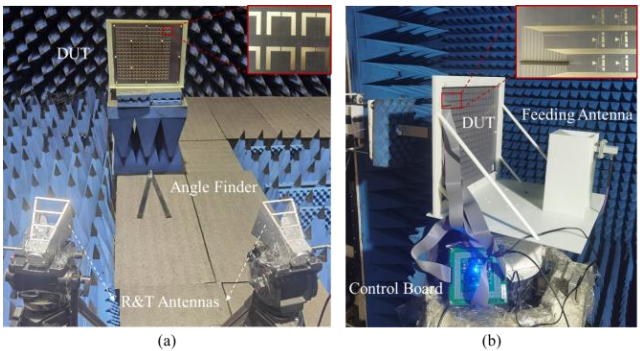

Fig. 12 Measured setup for (a) RCS reduction and (b) beam scanning performance.

### B    Radiation Performance

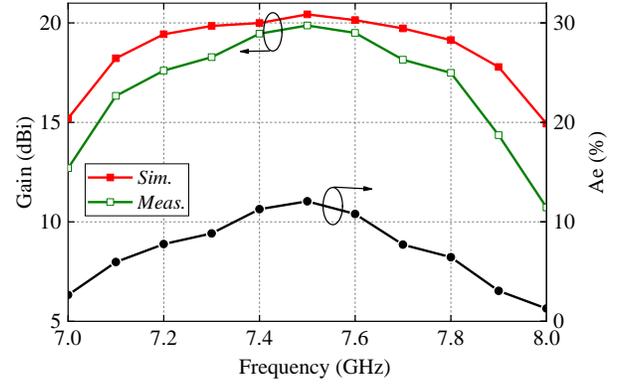

Fig. 13 Simulated and measured gain of the designed low RCS 1 bit RTA and its corresponding aperture efficiency.

The simulated and measured gain of the designed low RCS 1 bit RTA prototype at different frequencies is illustrated in Fig. 13. In the operating band, the realized gain is approximately 19.87 dBi, with a corresponding aperture efficiency of 12.1%. Compared to the simulation results, the measured gain of the prototype is reduced by approximately 0.6 dB, and the gain bandwidth is also slightly reduced. However, taking into account machining errors, test errors, and slight warping of the AFST layer, the results are considered acceptable.

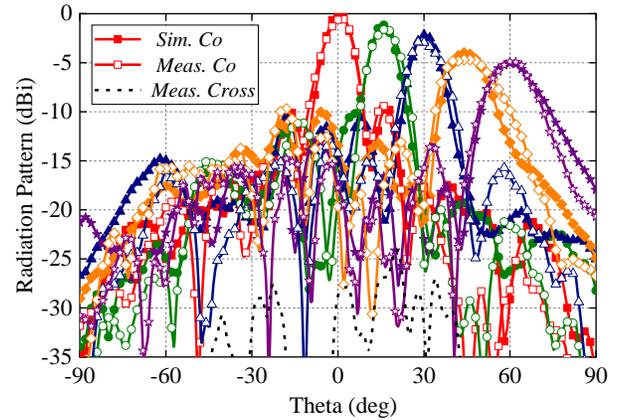

(a)

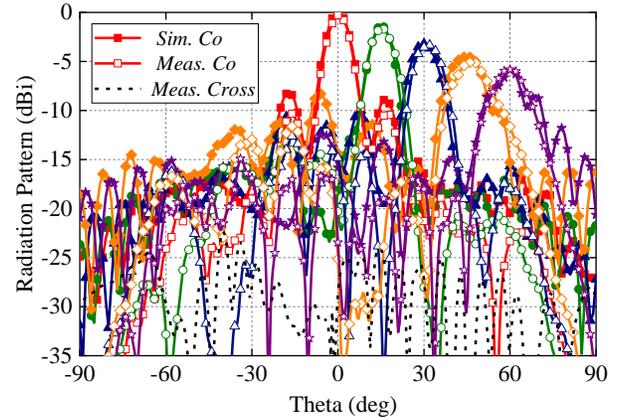

(b)

Fig. 14 Simulated and measured beam scanning performance of the designed low RCS 1 bit RTA, (a) E-plane, (b) H-plane.

The beam scanning performance of the low RCS 1-bit RTA

in E- and H- planes are shown in Fig. 14. It is evident that the beam scanning accurately reaches the expected direction within the range of 60°, and the scanning gain loss is about 5 dB. Additionally, the sidelobe level of the scanning beam is about -10 dB. This is mainly due to the small aperture of the prototype, which causes performance deterioration when the scanning angle is large due to edge effects. Additionally, the cross-polarization level is less than -25 dB in both principal planes. Overall, the measured results demonstrate that the proposed low RCS 1-bit RTA is capable of achieving 60° two-dimensional beam scanning while maintaining good high-gain radiation performance.

### C  RCS Reduction Performance

The RCS reduction performance of the prototype can be determined from the simulated results shown in Fig. 15. Fig. 15(a) displays the monostatic RCS with and without AFST. As shown, the integration of AFST on top of the RTA effectively reduces the out-of-band RCS. Specifically, the 10 dB RCS reduction bandwidth for low and high frequencies is 37% and 9.4%, respectively. Furthermore, Fig. 15(b) and (c) illustrate the simulated scattering patterns at low frequency of 4.5 GHz and high frequency of 11.2 GHz, respectively. It is apparent that the designed structure significantly reduces the out-of-band RCS compared to the RTA without AFST.

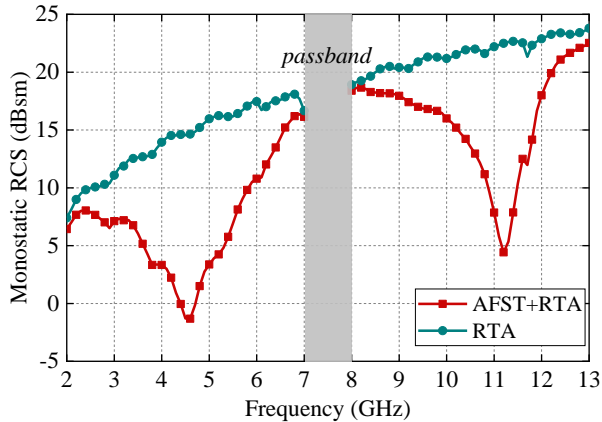

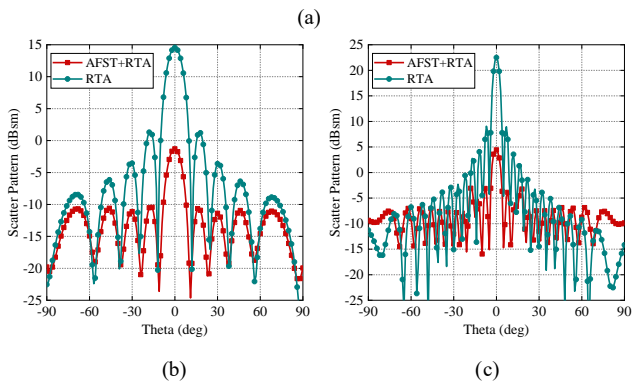

Fig. 15 Simulated RCS of the designed low RCS 1 bit RTA, (a) monostatic RCS, (b) scatter pattern at 4.5 GHz and (c) scatter pattern at 11.2 GHz.

On the other hand, the RCS reduction performance under oblique incidence is given in Fig. 16. In the case of 30° oblique incidence, the designed low RCS 1 bit RTA can maintain relatively stable RCS reduction performance. In addition, it should be pointed out that RCS reduction is mainly carried out for the incident wave with the same polarization of RTA, while the incident wave with cross polarization can be effectively reduced by simply constructing a wideband absorber, so it is not specifically involved in this work.

The simulated and measured monostatic RCS reduction performance under oblique incidence is shown in Fig. 16. It is observed that, in the case of 30° oblique incidence, the designed low RCS 1 bit RTA can maintain relatively stable RCS reduction performance. Furthermore, it should be noted that the RCS reduction primarily applies to the incident wave with the same polarization as the RTA. The incident wave with cross polarization can be effectively reduced by simply constructing a wideband absorber, so it is not specifically involved in this work.

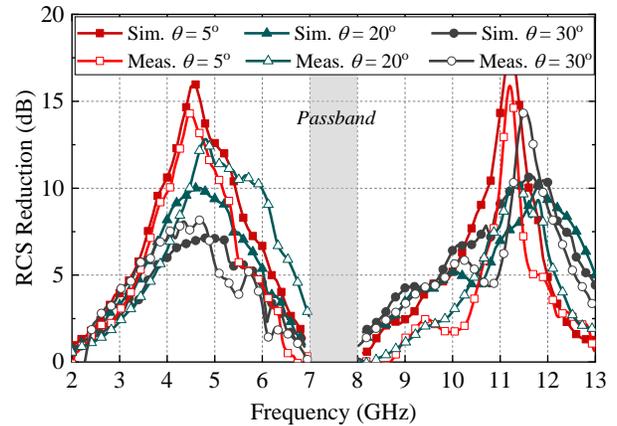

Fig. 16 Simulated and measured monostatic RCS with various incident angle.

Comparison of the proposed low-RCS 1-bit RTA with those reported in several previously published works reveals that the primary advantage of our work is the simultaneous implementation of in-band high gain radiation and beam scanning, along with out-of-band RCS reduction performance. Additionally, the designed structure has a low profile, simple processing, makes it a promising candidate for use in radar systems and other related fields.

## V. CONCLUSION

This paper presents a 1-bit electronically reconfigurable transmitarray with out-of-band scatter suppression, consisting of the AFST and RTA layers. The AFST layer uses the first three resonant modes of a bent metallic strip with a centrally loaded resistor to achieve out-of-band scattering suppression and in-band transmission performance. Meanwhile, the RTA layer adopts a receiver-transmitter structure with an active receiving dipole and a passive orthogonal transmitting dipole to achieve 1-bit transmission phase reconfiguration. Subsequently, a 16×16-element transmitarray prototype is designed and tested, revealing in-band high gain radiation and beam scanning capabilities, along with out-of-band RCS reduction.